\documentclass[twocolumn,amsmath,amssymb,floatfix,prb,showpacs,footinbib,superscriptaddress]{revtex4}
\usepackage{amsmath,amssymb,natbib,bm,graphicx,url,psfrag,times}
\usepackage[ansinew]{inputenc}

\newcommand{\be}{\begin{equation}}
\newcommand{\ee}{\end{equation}}
\newcommand{\bes}{\begin{equation}\begin{split}}
\newcommand{\ees}{\end{split}\end{equation}}

\newlength{\myVSpace}
\newcommand\xstrut{\raisebox{-.5\myVSpace}
  {\rule{0pt}{\myVSpace}}%
}

\newcommand{\vc}[1]{\mathbf{#1}}

\newcommand{\abs}[1]{\left|#1\right|}

\newcommand{\ket}[1]{\left|\, #1 \, \right\rangle}

\begin{document}
\title{Nonequilibrium charge-Kondo transport through negative-$U$ molecules}
\author{Jens Koch}
\affiliation{Department of Applied Physics, Yale University, New Haven, Connecticut 06520, USA}
\thanks{current address}
\affiliation{Institut f\"ur Theoretische Physik, Freie Universit\"at Berlin, Arnimallee 14, 14195 Berlin, Germany}
\author{Eran Sela}
\affiliation{Department of Condensed Matter Physics, Weizmann Institute for Science,  Rehovot 76100, Israel}
\author{Yuval Oreg}
\affiliation{Department of Condensed Matter Physics, Weizmann Institute for Science,  Rehovot 76100, Israel}
\author{Felix von Oppen}
\affiliation{Institut f\"ur Theoretische Physik, Freie Universit\"at Berlin, Arnimallee 14, 14195 Berlin, Germany}
\date{October 22, 2006}
\begin{abstract}
Low-temperature transport through molecules with effectively negative charging energy $U$ exhibits a charge-Kondo effect. We explore this regime analytically by establishing an exact mapping between the negative-$U$ and the positive-$U$ Anderson models, which is suitable for the description of nonequilibrium transport. We employ this mapping to demonstrate the intimate relation between nonequilibrium tranport in the spin-Kondo and charge-Kondo regimes, and derive analytical expressions for the nonlinear current-voltage chracteristics as well as the shot noise in the latter regime. Applying the mapping in the opposite direction, we elucidate the finding of super-Poissonian noise in the positive-$U$ Anderson model at high temperatures, by relating the correlations between spin flips to pair-tunneling processes at negative $U$.
\end{abstract}
\maketitle
\section{Introduction}
Recently, several theoretical studies have substantiated that transport through single molecules may realize the scenario of an Anderson model with \emph{negative charging energy} $U$.\cite{alexandrov1,alexandrov2,cornaglia1,cornaglia2,arrachea,koch6} 
The negative $U$ originates from the coupling of the electronic orbital $\varepsilon_d$ of the molecule to the molecular vibrations, by a mechanism originally introduced by Anderson in the context of amorphous semiconductors.\cite{anderson2} Assuming that real excitation of the molecular vibrations is energetically not allowed,
transport is described by an Anderson model 
\begin{align}\label{negU}
H=&\varepsilon_d \sum_\sigma d_{\sigma}^\dag d_{\sigma} + U n_{d\uparrow}n_{d\downarrow} + \sum_{ak\sigma} (\epsilon_{ak}- eV_a)c_{ak\sigma}^\dag c_{ak\sigma}\nonumber\\
& +\sum_{ak\sigma}\left(t_{ak} c_{ak\sigma}^\dag d_{\sigma} +  \text{h.c.}\right),
\end{align}
with an effectively attractive on-site interaction $U<0$. Tunneling of electrons between the molecule and the left (right) electrode with amplitude $t_{a=L,k}$ ($t_{a=R,k}$) is driven by the applied bias $V=V_L-V_R$. Since only energies in the vicinity of the Fermi energy are relevant to transport, we assume the hopping amplitudes $t_a$ independent of $k$, and a linear dispersion relation $\epsilon_{ak}$.

As a result of the attractive interaction, the system favors even electronic occupation numbers and may develop a degeneracy between two even-number charge states. At high temperatures, finite-bias transport in the vicinity of this degeneracy point is accomplished by tunneling of electron pairs. In Ref.~\onlinecite{koch6}, two of us have demonstrated that this pair-tunneling regime can be addressed analytically, and manifests itself in distinct features in the linear conductance as well as in the nonlinear current-voltage characteristic ($IV$). In the low-temperature regime $T<T_K$, the degeneracy between the two charge states induces electronic correlations and leads to a charge-Kondo effect.\cite{coleman} To date, studies of transport in this regime have focused on numerical calculations of the linear conductance.\cite{cornaglia1,cornaglia2,arrachea} 

The central goal of the present paper is to develop an \emph{analytical} treatment of {\em linear and nonlinear} transport in the charge-Kondo regime. We achieve this by establishing a one-to-one mapping between the negative-$U$ and the conventional (positive-$U$) Anderson model, which we term particle-hole/left-right (PHLR) tranformation. The particle-hole transformation forming the first part of this mapping was introduced by Iche and Zawadowski,\cite{iche} and further elucidated by Haldane.\cite{haldane} This mapping converts between spin and charge degrees of freedom of both the localized level and the leads. For this reason, this mapping by itself is not well suited for a description of nonequilibrium transport. We show that a subsequent left-right transformation eliminates this problem and the resulting combined mapping interchanges charge and spin degrees of freedom of the localized level only. We demonstrate that this allows for a transparent and analytical investigation of transport in the charge-Kondo regime based upon well-known results\cite{pustilnik2} for the conventional spin-Kondo effect.

New physical insight can also be gained by exploiting the PHLR transformation in the opposite direction, i.e.\ translating from the negative-$U$ to the conventional Anderson model. For negative-$U$, pair-tunneling processes make it natural to expect super-Poissonian current noise. The PHLR mapping then implies that Fano factors $F>1$ must also occur in the positive-$U$ Anderson model, where this result is surprising at first sight. We show that this noise enhancement can be traced back to the influence of a local Zeeman field (which is generated by the mapping). This Zeeman field induces correlations between consecutive spin-flip processes.

The presentation of our results is organized as follows. In Section \ref{sec:mapping} we introduce the PHLR mapping, which allows us to translate between positive and negative $U$, and serves as the central tool in our subsequent considerations. The conversion from negative to positive $U$ is exploited in Section \ref{pairnoise}, where we focus on the occurence of super-Poissonian noise at high temperatures outside the Kondo regime. The opposite mapping direction, i.e.\ from positive to negative $U$, is employed in Section \ref{sec:pos-neg} to obtain results for transport in the charge-Kondo regime. A summary and discussion of our results are given in Section \ref{sec:sum}. Some additional results on shot noise in the high-temperature regime of the negative-$U$ model, which are not necessary for an understanding of the main text, are relegated to an appendix. 

\section{Mapping between the negative-$U$ and positive-$U$ Anderson models\label{sec:mapping}}

The low-temperature physics of the positive-$U$ and the negative-$U$ Anderson models are intimately related: The former gives rise to the conventional spin-Kondo effect,\cite{kondo} the latter exhibits the charge-Kondo effect.\cite{coleman} The essential ingredients for the conventional Kondo effect are (i) a spin-degenerate localized orbital, and (ii) the $SU(2)$ spin symmetry (including spin-independent hopping amplitudes). By close analogy, the central requirements for the charge-Kondo effect are (i') degeneracy of two charge states which transform into each other under particle-hole transformation, and (ii') particle-hole symmetry of the entire system including the leads.

This close similarity suggests that a transformation interchanging spin and charge degrees of freedom and thus converting  between $SU(2)$ spin symmetry and particle-hole symmetry may serve as a tool for treating both Kondo effects on the same footing. Indeed, as first established by Iche and Zawadowski \cite{iche}, this is accomplished by a particle-hole transformation restricted to one spin direction which maps the charging energy $U$ into $-U$.  

The interchange of spin and charge degrees of freedom has several immediate consequences. Tuning the gate voltage in the negative-$U$ model away from the charge degeneracy point causes particle-hole symmetry breaking. This breaking of particle-hole symmetry is mapped into a breaking of SU(2) spin symmetry in the corresponding positive-$U$ model. Indeed, the detuning of the gate voltage from the charge degeneracy point at negative $U$ maps into a local Zeeman field
acting on the localized orbital at positive $U$. Thus, unlike the spin-Kondo effect which persists over a wide range of gate voltages, the charge Kondo effect fully develops only exactly at the charge degeneracy point. Similarly, the spin symmetry of the negative-$U$ model is mapped into particle-hole symmetry of the corresponding positive-$U$ model. In the absence of a real Zeeman field in the original negative-$U$ model, the positive-$U$ model is thus fixed to the particle-hole symmetric point $\varepsilon_d'=-U'/2$. (Here, we denote quantities belonging to the positive-$U$ model by primes.) More generally, a real Zeeman field only leads to breaking of the particle-hole symmetry after the mapping into a positive-$U$ model. The charge-Kondo effect is thus much less sensitive to (real) Zeeman fields than the spin-Kondo effect. 

Accordingly, the central results of the mapping can be summarized by the following ``dictionary",
 \begin{align}
 \varepsilon_d'=(U+B)/2, \quad U' = -U, \quad B'=2\varepsilon_d+U, \label{dictionary}
 \end{align}
where $B'$ denotes the local Zeeman field in the positive-$U$ model. In extension to Eq.~\eqref{negU}, we have also included a possible Zeeman field $B$ in the negative-$U$ model to emphasize the complete formal analogy between the two models. However, in the following we will restrict our considerations to zero Zeeman field in the original negative-$U$ model. We now turn to a more detailed derivation of these results, following References \onlinecite{iche} and \onlinecite{haldane}.

\subsubsection*{Particle-hole transformation}
The particle-hole (PH) transformation is carried out for one of the two spin directions, and without loss of generality we may choose $\sigma=\downarrow$. We define new annihilation and creation operators $\beta$, $\beta^\dag$ by
\begin{align}
\beta_{d\uparrow}&\equiv d_{\uparrow},            & \beta_{ak\uparrow}&\equiv c_{ak\uparrow},\\
\beta_{d\downarrow}&\equiv -d_{\downarrow}^\dag,   & \beta_{ak\downarrow}&\equiv c_{a(-k)\downarrow}^\dag.
\end{align}
It is straightforward to verify that the $\beta$ operators obey the usual anticommutation rules. We define the corresponding number operators by $\bar{n}_{d\sigma}\equiv \beta_{d\sigma}^\dag \beta_{d\sigma}$ and $\bar{n}_{ak\sigma}\equiv \beta_{ak\sigma}^\dag \beta_{ak\sigma}$. From this, one immediately infers that
\begin{align}
n_{d\uparrow}  &=\bar{n}_{d\uparrow} ,            & n_{ak\uparrow}&=\bar{n}_{ak\uparrow},\\
n_{d\downarrow}&=1-\bar{n}_{d\downarrow},         & n_{ak\downarrow}&=1-\bar{n}_{a(-k)\downarrow},
\end{align}
and, in particular
\begin{align}
U n_{d\uparrow}n_{d\downarrow} &= U \bar{n}_{d\uparrow} - U \bar{n}_{d\uparrow}\bar{n}_{d\downarrow}.
\end{align}
The latter equation explains the origin of the sign change of the charging energy, and the emergence of a Zeeman field. Altogether, the PH transformation maps the original Hamiltonian (up to an irrelevant additive constant) to
\begin{align}\label{ph-hamiltonian}\nonumber
&\bar{H} =  \frac{U}{2}\sum_\sigma \bar{n}_{d\sigma} - U\bar{n}_{d\uparrow}\bar{n}_{d\downarrow} +\sum_{ak\sigma}  \left(\epsilon_{ak}-a\sigma eV/2\right)\bar{n}_{ak\sigma} \\
&+ \sum_{ak\sigma}\left( t_{a} \beta_{ak\sigma}^\dag \beta_{d\sigma} + \text{h.c.} \right)
+ (\varepsilon_d+U/2)(\bar{n}_{d\uparrow} - \bar{n}_{d\downarrow}).
 \end{align}
Here and in the following, we assume symmetric voltage splitting between the leads, i.e.\ $V_a= aV/2$. [For easier notation, we identify $a=L,R$ with $a=+1,-1$.]
We have also chosen to measure $k$ with respect to the Fermi momentum $k_F$, so that $\epsilon_{a(-k)}=-\epsilon_{ak}$.
Analyzing the resulting expression for the Hamiltonian, one notes that the Anderson model with negative $U$ has been mapped into a  model with positive $U'=-U$, which includes an additional Zeeman field $B'$. 
However, we emphasize that the current through the system, originally given by
\be\label{origcur}
\langle I \rangle =\frac{e}{2}\left\langle\frac{d}{dt} \sum_{ak\sigma} an_{ak\sigma} \right\rangle,
\ee
now takes the form
\be\label{spincur}
\langle I \rangle   = \frac{e}{2}\left\langle\frac{d}{dt} \sum_{ak\sigma} a\sigma \bar{n}_{ak\sigma} \right\rangle.
\ee
The latter equation clearly demonstrates the crucial problem of the PH transformation when considering nonequilibrium transport: By the PH mapping, the charge current of the negative-$U$ model [Eq.~\eqref{origcur}] is turned into a \emph{spin} current in the conventional Anderson model [Eq.~\eqref{spincur}]. Moreover, finite voltages translate into Zeeman field gradients in the transformed Hamiltonian, see Eq.~\eqref{ph-hamiltonian}. As a consequence, the PH transformation by itself is not a well-suited tool for the analysis of nonlinear transport. 

\begingroup
\squeezetable
\begin{table*}
	\centering
    \begin{ruledtabular}
		\begin{tabular}{ccccccc}
		& dot operators & lead operators & dot state & charging energy & gate voltage & Zeeman energy\\
		&               &                &           &                 & $\rightarrow$Zeeman energy & $\rightarrow$gate voltage\\\hline
		\xstrut
		\parbox{2cm}{\textbf{negative-$U$}\\\textbf{model}}
		& $d_\uparrow$, $d_\downarrow$
		& $c_{ak\uparrow}$, $\qquad c_{Lk\downarrow}$, $\qquad c_{Rk\downarrow}$
		& $\ket{0}$, $\;\ket{\uparrow}$, $\;\ket{\downarrow}$, $\;\ket{\uparrow\downarrow}$
		& $U<0$
		& $\varepsilon_d$
		& $B$ \\\hline
		\xstrut
		\parbox{2cm}{\textbf{positive-$U$}\\\textbf{model}}
		& $d'_\uparrow$, $-{d'}^\dag_\downarrow$
		& $c'_{ak\uparrow}$, ${c'}^\dag_{R(-k)\downarrow}$, ${c'}^\dag_{L(-k)\downarrow}$
		& $\ket{\downarrow}'$, $\ket{\uparrow\downarrow}'$, $\ket{0}'$, $\ket{\uparrow}'$
		& $U'=-U>0$
		& $B'=2\varepsilon_d+U$
		& $\varepsilon_d'=(U+B)/2$
		\end{tabular}
		\caption{Dictionary for the PHLR mapping between the negative-$U$ model and the conventional Anderson model with additional Zeeman field. Symbols with (without) primes denote quantities in the postive-$U$ (negative-$U$) model.\label{dict}}
		\end{ruledtabular}
\end{table*}
\endgroup

\subsubsection*{Left-right transformation}

A solution is provided by a left-right (LR) transformation, performed in addition to the PH transformation. 
The LR transformation interchanges the roles of left and right leads for one spin direction. In the following, we establish this transformation, and show that for symmetric devices (i.e., symmetric molecule-lead coupling and symmetric voltage splitting) the combined PHLR transformation meets all requirements for a one-to-one mapping suitable for nonlinear transport.

For later convenience, we first carry out the LR transformation for general molecule-lead couplings, and only subsequently specialize to symmetric devices. The transformation consists of an interchange of left and right labels for the spin component previously affected by the particle-hole transformation, i.e.,
\begin{align}
d'_{\sigma}&\equiv \beta_{d\sigma},           &c'_{ak\uparrow}&\equiv \beta_{ak\uparrow}, \\
c'_{Lk\downarrow}&\equiv \beta_{Rk\downarrow}, & c'_{Rk\downarrow}&\equiv \beta_{Lk\downarrow}.
\end{align}
This transforms the  Hamiltonian $\bar{H}$ into
\begin{align}\label{mapped}
&H'= \frac{U}{2}\sum_\sigma n'_{d\sigma} - Un'_{d\uparrow}n'_{d\downarrow} +\sum_{ak\sigma} ( \epsilon_{ak}-a eV/2)n'_{ak\sigma}  \nonumber\\
&+ \sum_{ak \sigma}\left( t'_{a \sigma} {c'}_{ak\sigma}^\dag
d'_{ \sigma} +  \text{h.c.}\right) + (\varepsilon_d+U/2)(n'_{d
\uparrow} - n'_{d \downarrow}),
 \end{align}
where
\be\label{tunmat}
t'_{a\uparrow} = t_{a}, \quad t'_{L\downarrow}=t_{R}, \quad t'_{R\downarrow}=t_{L}.
\ee
As noted above, primes denote parameters of the positive-$U$ model, after the combined PHLR-transformation.
The crucial point of Eq.~\eqref{mapped} is that the bias now affects both spin components in the same way, and the current is expressed as
\be
\langle I \rangle = \frac{e}{2}\left\langle\frac{d}{dt} \sum_{ak\sigma} a n'_{ak\sigma} \right\rangle.
\ee
Through the combined PHLR mapping, the charge current of the negative-$U$ model [Eq.\ (\ref{origcur})] is thus mapped into a charge current for the positive-$U$ model.

\subsection{Asymmetric devices in the linear-response regime\label{sec:linresp}}

A difficulty arising for asymmetric molecule-lead couplings consists of the spin dependence of hopping within the positive-$U$ model, see Eq.\ \eqref{tunmat}. In this section, we first show that this complication drops out when restricting to the linear-response regime.\footnote{We thank A.\ Schiller for an instructive discussion of this point.} Specifically, we show that the conductance of the spin-dependent model $H'$, Eq.\ \eqref{mapped}, is identical to the conductance of an ordinary Anderson model 
\begin{align}\label{mapped2}
&H''= \frac{U}{2}\sum_\sigma n'_{d\sigma} - Un'_{d\uparrow}n'_{d\downarrow} +\sum_{ak\sigma} ( \epsilon_{ak}-a eV/2)n'_{ak\sigma}  \nonumber\\
&+ \sum_{ak \sigma}\left( t_{a} {c'}_{ak\sigma}^\dag
d'_{ \sigma} +  \text{h.c.}\right) + (\varepsilon_d+U/2)(n'_{d
\uparrow} - n'_{d \downarrow}),
 \end{align}
obtained from $H'$ by substituting Eq.~\eqref{tunmat} with the spin-independent prescription $t'_{a\sigma}=t_{a}$.

The calculation of the conductance is simplified by invoking a canonical transformation of the lead operators, henceforth referred to as Glazman-Raikh transformation for definiteness.\cite{glazman3} Conventionally, i.e.\ for spin-independent tunneling described by $H''$, this transformation converts the left and right lead electrons into two channels $\psi_{ik\sigma}$ ($i=1,2$)  according to
\be
\left(
\begin{array}{l}
\psi_{1k\sigma}\\\psi_{2k\sigma}	
\end{array}
\right)
=
\left(
\begin{array}{cc}
c & s\\
-s & c
\end{array}
\right)
\left(
\begin{array}{l}
c_{Rk\sigma}\\ c_{Lk\sigma}	
\end{array}
\right),
\ee
where $(c,s)=\frac{(t_{R},t_{L})}{\sqrt{\sum_a t_{a}^2}}$. This decouples the channel $i=2$ from the localized level, and the tunneling term is transformed into 
\be\label{trans-tun}
\sum_{k\sigma}[\tilde{t}\,\psi_{1k\sigma}d_\sigma + \text{h.c.}],
\ee
 where $\tilde{t}=\sqrt{\sum_a t_{a}^2}$. As a consequence, the current operator can be expressed in terms of the decoupled channel only,
\be\label{cur-op}
I=-\frac{i}{\hbar}\sum_{k\sigma}\frac{1}{\tilde{t}}\left[t_{L} t_{R} \psi_{2k\sigma}^\dag d_\sigma -\text{h.c.}\right].
\ee
Turning to the case of spin-dependent hopping described by $H'$, we find that the extension of the Glazman-Raikh transformation is straightforward, and merely amounts to the introduction of spin dependences $t_{a}\to t_{a\sigma}$, $\tilde{t}\to\tilde{t}_\sigma$. The crucial point in our proof is that for the spin-dependent tunneling matrix elements given by Eq.\ \eqref{tunmat}, the expressions $\tilde{t}_\sigma=\sum_{a}t_{a\sigma}^2$ and $t_{L\sigma}t_{R\sigma}$ in fact remain \emph{spin independent}. As a result, the Glazman-Raikh transformation of $H'$ and $H''$ leads to the same tunneling terms [Eq.~\eqref{trans-tun}] and current operators [Eq.\ \eqref{cur-op}], respectively.
Employing the Kubo formalism,\cite{kubo} the conductance is expressed in terms of the zero-frequency current-current correlation function in equilibrium. Since the Glazman-Raikh transformation of $H'$ and $H''$ leads to the same expressions for the Hamiltonian and current operators, the conductances are identical.

We emphasize that these arguments do \emph{not} apply to the nonequilibrium case. For finite bias, the different chemical potentials of the left and right lead prohibit the channel decoupling via the Glazman-Raikh transformation.

\subsection{Symmetric devices out of equilibrium}
\begin{figure}
	\centering
  \psfrag{a}[c][][0.8]{$\ket{\uparrow},\ket{\downarrow}\, \varepsilon_d$}
  \psfrag{b}[c][][0.8]{$\ket{0}$}
  \psfrag{c}[c][][0.8]{$\ket{\uparrow\downarrow}\, \varepsilon_d+U/2$}
  \psfrag{d}[c][][0.8]{$\ket{\uparrow\downarrow}'\, \varepsilon_d'+U'/2$}
  \psfrag{e}[c][][0.8]{$\ket{0}'$}
  \psfrag{f}[c][][0.8]{$\ket{\uparrow}'\, \varepsilon_d'+B'/2$}
  \psfrag{g}[c][][0.8]{$\ket{\downarrow}'\, \varepsilon_d'-B'/2$}
  \psfrag{v}[c][][0.8]{$eV/2$}
  \psfrag{w}[c][][0.8]{$-eV/2$}
		\includegraphics[width=1\columnwidth]{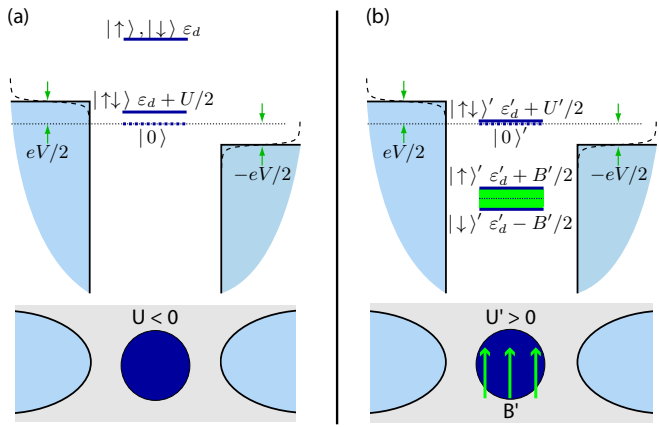}
	\caption{Illustration of the mapping between the negative-$U$ model (a) and the conventional Anderson model (b) with additional local Zeeman field $B'$. All level energies are given as energies per particle.\cite{fno1}  The horizontal dashed line marks the zero-bias Fermi energy of the leads. The one-particle energy $\varepsilon_d$ is required for adding a single electron to the neutral molecule.\label{fig:map1}}
\end{figure}

The spin dependence of the tunneling matrix elements Eq.~\eqref{tunmat} drops out for symmetric couplings, i.e.\  when $t_L=t_R$ we recover an ordinary Anderson model after the PHLR transformation. 
As a result, for symmetric devices we have established a one-to-one mapping between transport in the negative-$U$ model and transport in a conventional Anderson model, which remains valid {\it beyond} the linear response.
The conventional Anderson model contains an additional local Zeeman field, whose magnitude is determined by the gate-voltage detuning from the degeneracy point in the negative-$U$ model. In addition, spin symmetry of the negative-$U$ model enforces particle-hole symmetry of the positive-$U$ model, which is thereby fixed to the symmetric point $2\varepsilon_d'+U'=0$. An illustration of the resulting configuration is provided in Fig.~\ref{fig:map1}. 
The details of the mapping are  summarized by the ``dictionary" in Table \ref{dict}.

\section{From negative to positive $U$: Consequences of a local Zeeman field\label{pairnoise}}

While the principal goal of this paper consists of applying the PHLR mapping to study transport in the negative-$U$ model at low temperatures, it is interesting to note that valuable insight can also be gained by employing the mapping in the opposite direction. Specifically, in this section we investigate the current noise for the positive-$U$ model with local Zeeman field at high temperatures $T\gg T_K$.\footnote{Here, we restrict our calculations to the leading order in the tunneling between molecule and leads. The inclusion of higher order terms, including inelastic contributions,\cite{kaminski} is an interesting open problem and beyond the scope of the present paper.}

\begin{figure}
	\centering
	  \psfrag{c}[][][0.9]{\raisebox{0.1cm}{$F$}}
  \psfrag{b}[][][0.9]{\raisebox{0.2cm}{$eV\; (10^{-3}\abs{U})$}}
  \psfrag{a}[][][0.9]{\raisebox{0.2cm}{$\varepsilon_d+\abs{U}/2\; (10^{-3}\abs{U})$}}
		\includegraphics[width=0.95\columnwidth]{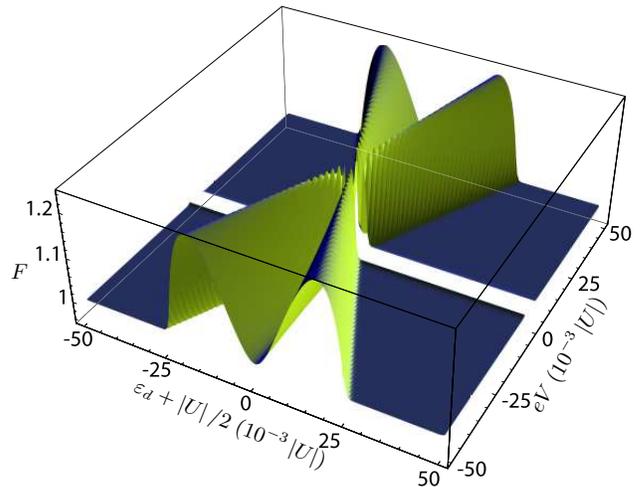}
	\caption{Fano factor as a function of bias and gate voltage for a symmetric junction with $\Gamma_L=\Gamma_R=k_BT$, and with $U=-2\times10^3k_BT$. Super-Poissonian noise reflects the bunching of electrons in pair-tunneling processes.\label{fig:fano}}
\end{figure}

Within the negative-$U$ model, transport in the high-temperature regime is accomplished by tunneling of electron pairs.\cite{koch6} Thus, it is natural to expect super-Poissonian zero-frequency noise $S$ noise due to electron pairing. A convenient measure of the effectively transferred charge is given by the Fano factor $F=S/2e|\langle I\rangle|$, as demonstrated for Cooper pairs in superconductors,\cite{jehl,muzykantskii} and quasiparticles with fractional charge in the quantum Hall regime.\cite{kane,depicciotto,saminadayar} 
In Fig.~\ref{fig:fano}, we present our numerical results for the negative-$U$ model, obtained via a generalized version of the technique developed by Korotkov.\cite{korotkov} (This formalism is dicussed in Ref.~\onlinecite{koch7}, and the computation for the pair-tunneling regime is detailed in Appendix \ref{app:noise}.)\footnote{During preparation of this manuscript, we became aware of related work on shot noise in the high-temperature limit of the negative-$U$ Anderson model by M.-J.\ Hwang, M.-S.\ Choi, and R.\ Lopez, cond-mat/0608599.} The plot depicts the Fano factor of a symmetric device at negative $U$, as a function of gate and bias voltage.
As expected, the transfer of electrons in pairs causes super-Poissonian noise $F>1$ in the regime dominated by pair tunneling. The fact that the value $F=2$, naively expected for electron pairs, is \emph{not} reached, is explained by the coexistence various pair-tunneling processes as well as cotunneling of single electrons. The Fano factor for a symmetric device typically reaches its maximum in the vicinity of the crossover between the cotunneling and pair-tunneling regimes. In this region, the phase space for pair tunneling onto the molecule ($\sim eV/2-\varepsilon_d-U/2$) is large compared to the  phase space for tunneling off the molecule ($\sim \varepsilon_d+U/2+eV/2\approx0$), or vice versa. As a result, one junction dominates the pair transport with a Fano factor close to 2. A more detailed understanding of this behavior may be obtained by an analysis of the different processes relevant for the charge transfer. This is demonstrated in Appendix \ref{app:noise}.

We now turn to the central point of employing the PHLR transformation, and map our negative-$U$ results to the conventional Anderson model with local Zeeman field. It is straightforward to obtain the corresponding expressions for the linear conductance and the nonlinear $IV$ as a function of the Zeeman field. Remarkably, beyond this the PHLR transformation leads to the conclusion that the current noise becomes super-Poissonian in the conventional Anderson model. This prediction comes as a surprise, since charge transfer at positive $U$ is essentially comprised of one-particle processes.

\begin{figure}
	\centering
		\includegraphics[width=1\columnwidth]{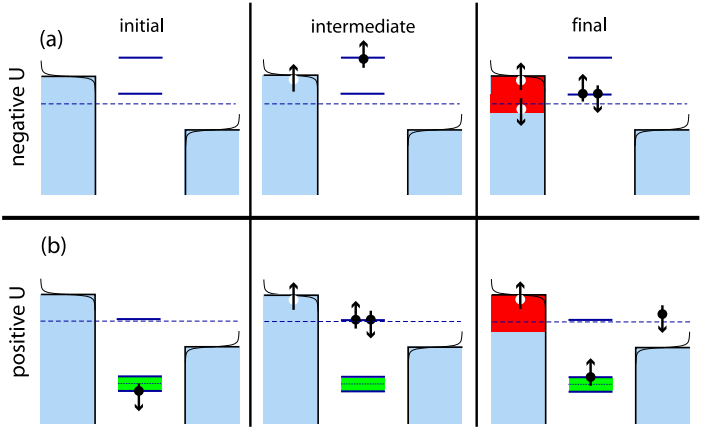}
	\caption{(Color online) Relation between (a) the unidirectional pair-tunneling process for negative $U$ and (b) the inelastic spin-flip in the conventional Anderson model with local Zeeman field, as obtained by the PHLR transformation. The phase space available for the spin-up electron in the initial state is marked in red (dark gray). \label{fig:map2}}
\end{figure}
The noise enhancement in the positive-$U$ model can be traced back to correlations induced by the local Zeeman field. Specifically, the origin of the super-Poissonian noise lies in the translation of the unidirectional pair tunneling [i.e.\ paired electrons are transferred across the same junction] into the language of the positive-$U$ Anderson model, see Fig.~\ref{fig:map2}. Applying the PHLR transformation, one finds that the two consecutive pair processes correspond to two inelastic cotunneling processes: The first process flips the spin from down to up, the second process flips it back to $\sigma=\downarrow$. In both cases, one electron is transferred from the left to the right lead. The crucial point is that these two processes are correlated whenever the Zeeman field is nonzero. For $B'>0$, a spin flip from $\sigma=\downarrow$ to $\sigma=\uparrow$ requires (twice) the Zeeman energy. The same energy is released for the opposite spin flip. As a result, the phase-space involved in these inelastic processes causes one spin flip to be faster than the other one. The difference of the rates becomes maximal when the Zeeman field becomes as large as the bias voltage, $B'\approx eV$.  These correlations cause the enhancement of noise beyond the Poissonian limit.

\section{From positive to negative $U$: Analytical study of the charge-Kondo effect\label{sec:pos-neg}}
In this section, we employ the PHLR mapping to study transport in the negative-$U$ model, both in the poor-man-scaling regime and the fully developed Kondo regime.

\subsection{The onset of the charge-Kondo effect: Logarithmic corrections}
The degeneracy between the two charge states $n=0$ and $n=2$ in the negative-$U$ Anderson model results in Kondo correlations at low temperatures.\cite{coleman} For temperatures $T\gg T_K$, the development of these correlations is expected to be signalled by logarithmic corrections to the leading-order perturbative results. Here, we exploit the PHLR transformation to extract the logarithms for the negative-$U$ case from well-known results in the context of the conventional Anderson model, see e.g.\ Ref.\ \onlinecite{pustilnik2}.

Specifically, we employ the following chain of transformations. Our starting point is the negative-$U$ Anderson model. Applying the PHLR transformation, we obtain a conventional Anderson model with local Zeeman field, Eq.~\eqref{mapped}. Then, a Schrieffer-Wolff transformation is performed.\cite{schrieffer} Up to an irrelevant renormalization of the level energy $\varepsilon_d$, the mapping results in $H'=H_0+H_\text{J}+H_\text{Z}$ with 
\begin{align}
H_0 =& \sum_{ak\sigma}( \epsilon_{ak} - aeV/2) c_{ak\sigma}^\dag c_{ak\sigma},\\\label{exch}
H_\text{J}=&\sum_{aa'kk'} J_{aa'kk'}\vc{s}_{aa'kk'}\cdot \vc{S},\\
H_\text{Z} =& B'S^z\label{eq:zim}.
\end{align}
Here, the exchange constant $J$ and the lead electron spin operator $\vc{s}$ are given by
\begin{align} \nonumber
J_{aa'kk'} =t_{a}'{t'}_{a'}&\bigg[ \frac{1}{\epsilon_{ak}-\varepsilon_d'}+\frac{1}{\epsilon_{a'k'}-\varepsilon_d'}+\frac{1}{U'+\varepsilon_d'-\epsilon_{ak}}\\
&+\frac{1}{U'+\varepsilon_d'-\epsilon_{a'k'}}\bigg],
\end{align}
and
\begin{align}
s_{aa'kk'}^z &= (c_{ak\uparrow}^\dag c_{a'k'\uparrow} - c_{ak\downarrow}^\dag c_{a'k'\downarrow})/2,\\
s_{aa'kk'}^+&=c_{ak\uparrow}^\dag c_{a'k'\downarrow},\qquad s_{aa'kk'}^-=c_{ak\downarrow}^\dag c_{a'k'\uparrow}. 
\end{align}
Note that the usual potential scattering term in $H'$ vanishes due to the particle-hole symmetry of the positive-$U$ model. As usual, the exchange is approximated by its Fermi-energy value, $J_{aa'kk'}\to J_{aa'}=-8t_a t_{a'}/U$ within a band of width $D$ and zero outside. In the linear-response regime, the Hamiltonian may be further simplified by diagonalizing $\mathsf{J}=(J_{aa'})$ with eigenvalues $0$ and $J=-(8/U)\sum_a \abs{t_a}^2$. 
In terms of the eigenvectors of $\mathsf{J}$, this results in the one-channel problem,\cite{pustilnik2}
\be\label{eq:tk}
H_K = \sum_i \sum_{k\sigma} \epsilon_{k} \psi_{ik\sigma}^\dag\psi_{ik\sigma}  +  J\vc{s}_1\cdot \vc{S} + B'S^z.
\ee
Alternatively, this can be derived by applying the Glazman-Raikh transformation\cite{glazman3} to the Anderson Hamiltonian before carrying out the Schrieffer-Wolff transformation.

After the mapping of the negative-$U$ problem to the Kondo Hamiltonian, we are now ready to translate results from the conventional spin-Kondo effect to the charge-Kondo scenario of negative $U$. In particular, a perturbative treatment of the exchange coupling leads to the well-known logarithmic terms $\sim J^2\ln (D/\omega)$ in the amplitude for a transition $\ket{ks,\sigma}\to\ket{k's',\sigma'}$.\cite{kondo,pustilnik2} Summing up the leading logarithmic contributions to all orders, or using  poor-man's scaling alternatively,\cite{anderson_pms} the exchange coupling is renormalized as $\nu J\to 1/\ln (T/T_K)$, where $T_K = D \exp[ -1/\nu J ]$ denotes the Kondo temperature, and $\nu$ the density of states of the leads. Employing the Kubo formula, the conductance for vanishing Zeeman field $B'=2\varepsilon_d+U=0$ is found to be\cite{pustilnik2}
\be\label{cond-t}
G=\frac{2e^2}{h}\frac{4\Gamma_L\Gamma_R}{(\Gamma_L+\Gamma_R)^2}\frac{3\pi^2}{16}\frac{1}{\ln^2(T/T_K)}.
\ee
Note that the leading order correctly reproduces the rate-equations result for the pair-tunneling peak: For $T\gg T_K$ we have $\ln (T/T_K)\approx 1/\nu J$, so that
\be
G\approx 24e^2 \Gamma_L\Gamma_R/U^2h,
\ee
in agreement with results in Ref.~\onlinecite{koch6}.
For decreasing temperature, the developing Kondo correlations cause a slow logarithmic increase of the peak height value. 

Additional corrections affect the tails of the conductance peak. In the language of the positive-$U$ Anderson model, this implies considering the situation of a large Zeeman field $B'=2\varepsilon_d+U\gg k_BT$. For this limit, a relation similar to Eq.~\eqref{cond-t} can be derived,\cite{pustilnik2} namely
\be
G=\frac{2e^2}{h}\frac{4\Gamma_L\Gamma_R}{(\Gamma_L+\Gamma_R)^2}\frac{\pi^2}{16}\frac{1}{\ln^2[(2\varepsilon_d+U)/T_K]}.
\ee
This captures the logarithmic corrections to the cotunneling tails of the conductance peak in the negative-$U$ model.

\subsection{The fully developed charge-Kondo regime}
We now turn to the investigation of the fully developed Kondo regime in the low-temperature limit, $T\ll T_K$, and employ the PHLR transformation to translate between the conventional spin-Kondo effect and the charge-Kondo effect. Our considerations are based on Nozi\`{e}res' Fermi-liquid description of the Kondo fixed-point Hamiltonian,\cite{nozieres1,nozieres2}  given by
\begin{align}
\label{eq:hfp} 
H =& \sum_{k \sigma} \xi_k \psi^\dagger_{k \sigma}\psi_{k \sigma}\nonumber \\
& 
-\frac{1}{\pi \nu T_K}\sum_{k,k' \sigma} \bigl( \alpha
\frac{\xi_k+\xi_{k'}}{2}+\gamma B' \sigma  \bigr)\psi^\dagger_{k
\sigma} \psi_{k' \sigma} \nonumber \\
&
+\frac{\beta}{\pi \nu^2 T_K}\sum_{k_1,k_2,k_3,k_4}
\psi^\dagger_{k_1 \uparrow} \psi_{k_2 \uparrow} \psi^\dagger_{k_3
\downarrow} \psi_{k_4 \downarrow},
\end{align}
where $\psi_{k\sigma}$ annihilates a quasiparticle with energy $\xi_k = v_F k$ and spin $\sigma$. The $\alpha$-term describes the quasiparticle scattering at the impurity site, the $\beta$-term accounts for the induced interaction between quasiparticles. As shown by Nozi\`{e}res, the fact that the Kondo resonance is floating on the Fermi sea, fixes the parameter ratio $\alpha/\beta$ to unity.\cite{nozieres1,nozieres2} The additional $\gamma$-term describes the effect of a local Zeeman field $B'$ acting on the impurity (as generated by the PHLR mapping). We note that the coefficients $\alpha$ and $\gamma$ take on the values $\alpha=\gamma=1$, when we interpret $T_K$ in Eq.\ (\ref{eq:hfp}) to be the exact Kondo temperature, as determined e.g., from the exact Bethe-ansatz solution,\cite{Tsvelick83} i.e., $T_K=T_K^{ba} = \frac{(2 U\Gamma)^{1/2}}{\pi} e^{-\frac{\pi U}{8 \Gamma}}$.

\subsubsection{Linear conductance}

Within the Fermi-liquid description, the linear conductance
may be expressed as a function of the scattering phase shift.
As shown by Nozi\`{e}res in Ref.~\onlinecite{nozieres1}, this phase shift depends both on the energy
of the incoming particle and on the quasiparticle distribution, and can be expanded as 
\begin{equation}
\delta_{\sigma} =\frac{\pi}{2}+ \frac{\alpha \epsilon}{ T_K} - \frac{\beta
n_{\bar{\sigma}}}{\nu T_K} 
+\frac{\gamma \sigma B' }{T_K}, \end{equation}
where we denote $\bar\uparrow = \downarrow$, $\bar{\downarrow}=\uparrow$.
The leading-order phase shift
$\pi/2$ makes the molecule a perfectly open
channel in the Kondo regime. From the phase shift we can determine the transmission and reflection amplitudes from left to right (from right to left), $t_\sigma$ and $r_\sigma$ ($t_\sigma'$ and $r_\sigma'$), respectively.
Indeed, using the Glazman-Raikh
transformation,\cite{glazman3} one obtains the corresponding scattering matrix
\begin{align}
\nonumber
\mathsf{S}_\sigma(\epsilon)&=
\left(%
\begin{array}{cc}
  r_\sigma & t'_\sigma \\
  t_\sigma & r'_\sigma \\
\end{array}%
\right) \\
&= \left(%
\begin{array}{cc}
  c & -s \\
  s & c \\
\end{array}%
\right)  \left(%
\begin{array}{cc}
  e^{2 i \delta_\sigma} & 0 \\
  0 & 1 \\
\end{array}%
\right)
\left(%
\begin{array}{cc}
  c & s \\
  -s & c \\
\end{array}%
\right).
\label{eq:Sm} 
\end{align}
This results in a transmission coefficient
\be
\abs{t_\sigma}^2 =
\frac{4\abs{t_L}^2\abs{t_R}^2}{( |t_L|^2+|t_R|^2 )^2}\sin^2(\delta_\sigma).
\ee
 At $T=0$, the linear
conductance is not expected to depend on $\alpha$ or $\beta$, 
since the quasi-particles have energy $\epsilon=0$ in linear response, and $n_{\bar{\sigma}}=0$ in the absence of 
a global magnetic field. As a result, the Kondo-regime conductance of the Anderson model with local Zeeman field $B' \ll T_K$ is given by
\be
G'=\frac{e^2}{h} \sum_\sigma |t_\sigma|^2=\frac{2e^2}{h}\frac{4\Gamma_L\Gamma_R}{( \Gamma_L+\Gamma_R )^2}
\left[1-(\gamma B'/T_K)^2 \right].
\ee
Exploiting the identity of linear conductances for the positive and negative-$U$ models, see Section \ref{sec:linresp}, we conclude that the charge-Kondo effect in the negative-$U$ model leads to the conductance
\be
G=\frac{2e^2}{h}\frac{4\Gamma_L\Gamma_R}{( \Gamma_L+\Gamma_R )^2}\left[1-\left(\frac{\gamma}{T_K} [2\varepsilon_d+U]\right)^2 \right].
\ee
Here, the correspondence between Zeeman field (at positive $U$) and gate voltage (at negative $U$) is directly reflected in a departure from the unitary limit as soon as the negative-$U$ system is tuned away from the charge-degeneracy point.

\subsubsection{Nonlinear current-voltage characteristics}

For the special case of symmetric junctions, the PHLR transformation allows us to go beyond linear response. 
For nonzero bias, quasiparticles with finite energy are involved, and hence
the $\alpha$- and $\beta$-terms in the effective Hamiltonian, Eq.~\eqref{eq:hfp}, become relevant. These terms describe the weak scattering off the spin-singlet state as well as the induced interaction between quasiparticles, and cause a reduction of the unitary current $I_u=2e^2/h\,V$ due to backscattering events. The total current may thus be written in the form $I=I_u-I_b$.
In order to extract the backscattering contributions from the effective Hamiltonian, we relate the quasiparticle states $\psi_{k\sigma}$ to left-movers and right-movers, $L_{k\sigma}$ and $R_{k\sigma}$, respectively. For a symmetric device, this relation is given by $\psi_{k\sigma}=\frac{1}{\sqrt{2}}(L_{k\sigma}+R_{k\sigma})$.\footnote{We emphasize that the $\psi_{k\sigma}$ states denote \emph{scattering} states with incoming wave vector $k$. They are \emph{not} momentum eigenstates and hence involve both left-moving and right-moving contributions.} Importantly, left-movers originate from the right lead and right-movers from the left lead. This fact allows one to account for the nonequilibrium situation at finite bias by identifying the distribution of right-movers (left-movers) with the Fermi distribution of the left (right) lead, i.e.\ $f_{L(R)}(\epsilon)$. Substituting the LR decomposition into the effective Hamiltonian, one finds that the elastic term ($\sim\alpha$) and the inelastic term ($\sim\beta$) generate backscattering transitions which turn left-movers into right-movers, and vice versa. Since these contributions act as weak perturbations, one may evaluate the backscattering current by summing the corresponding backscattering rates obtained via Fermi's golden rule. 

Following Ref.\ \onlinecite{sela} and including the additional local Zeeman term results in the total current
\begin{equation} \label{eq:res} 
I=
\frac{2e^2}{h}V \left[1-(\gamma B'/T_K)^2- \frac{\alpha^2+5
\beta^2}{12} \left( \frac{V}{T_K} \right)^2 \right].
\end{equation}
Due to the onset of backscattering at finite energies, the current is \emph{reduced} with increasing bias voltage. The breaking of spin-symmetry by the local Zeeman field leads to an additional reduction $\sim (B')^2$, as required by symmetry. Now, we return to the charge-Kondo effect in the negative-$U$ model by applying the PHLR transformation.  
The resulting current close to the unitary limit is given by
\begin{equation} I=
\frac{2e^2}{h}V \left[1-(\gamma [2\varepsilon_d+U]/T_K)^2- \frac{\alpha^2+5
\beta^2}{12} \left( \frac{V}{T_K} \right)^2 \right],
\end{equation}
revealing the current reduction due to a gate detuning from the charge degeneracy point.

\subsubsection{Shot noise}

We finally turn to a discussion of shot noise in the Kondo regime, which we have recently investigated for the conventional Anderson model in Ref.~\onlinecite{sela}. The PHLR transformation allows us to transfer our central results for the conventional  Anderson model  to the negative-$U$ model. 

As demonstrated in the previous subsection, close to $T=0$ and for small bias voltages, the current is nearly unitary, and backscattering events are rare. In this scenario, a sensible definition of the Fano factor does not involve the transmitted current (which would yield $F\simeq 0$), but rather the backscattering current $I_b$,
\begin{equation}
\label{eq:qef} F=
 {S}/{2e|I_b|}.
\end{equation}
The backscattering current consists of a competition between single-particle and pair backscattering
processes.
Remarkably, for vanishing local Zeeman field and the conventional positive-$U$ Anderson model, we have shown that the Fano factor is super-Poissonian, $F=5/3$, due to the pair backscattering.\cite{sela} It is the enhanced phase space for pair scattering events which renders their contribution dominant, despite the fact that they constitute only one of the seven relevant processes.\cite{sela} The inclusion of an additional local Zeeman field is straightforward, and we obtain
\be\label{newfano}
F=\frac{2(B'/T_K)^2+\frac{5}{3}(V/T_K)^2}{2(B'/T_K)^2+(V/T_K)^2}.
\ee
(Here, we have exploited the identity of the coefficients $\alpha$ and $\gamma$ following from the Bethe ansatz.\cite{Tsvelick83}) Hence, for $B'=0$ we reproduce the result $F=5/3$, while
in the opposite limit of large Zeeman fields $B'\gg V$, single-particle backscattering at the singlet dominates and the Fano factor is Poissonian, $F=1$.

Applying the PHLR mapping now yields the corresponding Fano factor for the negative-$U$ model,
\be
F=\frac{2([2\varepsilon_d+U]/T_K)^2+\frac{5}{3}(V/T_K)^2}{2([2\varepsilon_d+U]/T_K)^2+(V/T_K)^2}.
\ee
The Fano-factor enhancement to $5/3$ only dominates the immediate vicinity of the charge-degeneracy point. A detuning from this point results in a suppression of the Fano factor, and Poissonian noise is recovered as soon as the detuning is large compared to the bias voltage.

\section{Summary\label{sec:sum}}

By an extension of the particle-hole transformation due to Iche and Zawadowski,\cite{iche} we have established a one-to-one mapping between the negative-$U$ and the positive-$U$ Anderson models appropriate for the situation of nonequilibrium transport. This mapping transforms spin degrees to charge degrees of freedom, converts between $SU(2)$ spin symmetry and particle-hole symmetry, and is thus ideally suited to the investigation of the spin- and charge-Kondo effects. In the case of symmetric devices, the mapping leads to a conventional Anderson model with spin-independent hopping terms. For this scenario, we have demonstrated the usefulness of the mapping by deriving analytical expressions for nonlinear transport in the negative-$U$ model, both in the poor-man-scaling and the fully developed Kondo regimes. 

Applying the mapping in the opposite direction, we have shown that an additional Zeeman field causes correlations in the high-temperature transport through a positive-$U$ device. These correlations are reflected in super-Poissonian noise, and can be directly related to pair-tunneling processes at negative $U$ for which enhanced Fano factors occur naturally.

In closing, we emphasize that the nonequilibrium physics in the charge-Kondo regime  remains a partially open question for asymmetric molecule-lead couplings. Specifically, for asymmetric couplings between the molecule and the leads, the hopping terms in the corresponding positive-$U$ Anderson model become spin dependent, a situation not commonly addressed in the Kondo literature to date. 

\begin{acknowledgments}
We would like to thank A.\ Schiller and M.\ Raikh for valuable discussions. Two of us (FvO, JK) gratefully acknowledge hospitality by the Weizmann Institute, made possible by the EU - Transnational Access program (RITA-CT-2003-506095). This work was supported in part by the DFG through Sfb 658 and Spp 1243 (FvO), Studienstiftung des dt.~Volkes and Yale University via a Quantum Information and Mesoscopic Physics Fellowship (JK), DIP (FvO and YO), as well as ISF and BSF (YO).
\end{acknowledgments}

\appendix

\section{High-temperature noise calculations\label{app:noise}}
\begin{figure}[h]
	\centering
		\includegraphics[width=0.8\columnwidth]{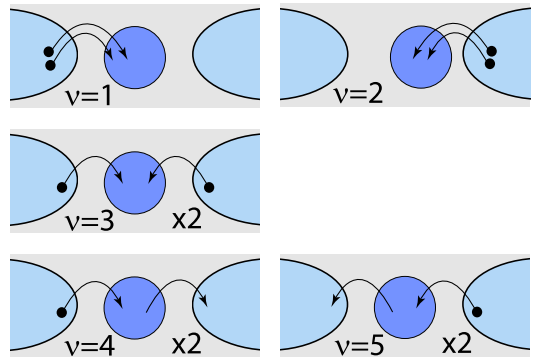}
	\caption{Elementary processes for the transitions $\ket{0}\to\ket{2}$ and $\ket{0}\to\ket{0}$. The label ``$\times2$" signals an additional factor of 2 due to two incoherent spin contributions. The diagrams for the processes corresponding to the transitions $\ket{2}\to\ket{0}$ and $\ket{2}\to\ket{2}$ are obtained by reversing the direction of all arrows.\label{fig:processes}}
\end{figure}

Our calculations of the current shot noise for pair tunneling are carried out by means of a generalized version of Korotkov's formalism,\cite{korotkov} which is detailed in Ref.\ \onlinecite{koch7}. This method is based on a decomposition of the time-dependent current into discrete contributions from individual tunneling processes.
The relevant processes labelled by $\nu=1,\ldots,5$ are identified in Fig.\ \ref{fig:processes}.  In addition to the transition rates, the formalism requires as input the charge $s_{if,\nu}^a$ (in units of e), transferred across junction $a$ for a process of type $\nu$ with initial and final state $i$ and $f$, respectively. For completeness, these charge-transfer coefficients are summarized in Table \ref{tab}. Note that in distinction to Ref.\ \onlinecite{koch7}, the absolute coefficients obtain values larger than unity due to pair tunneling.
\begin{table}
	\centering
		\begin{tabular}{r|c|c||r|c|c}
		$s_{02,\nu}^a$ & $a=L$ & $a=R$ & $s_{20,\nu}^a$ & $a=L$ & $a=R$ \\\hline
		$\nu=1$        & 2 & 0 & $\nu=1$        & $-2$ & 0\\
		$2$ & $0$ & $-2$ & $2$ & $0$ & $2$\\
		$3$ & $1$ & $-1$ & $3$ & $-1$ & $1$\\\hline\hline
		$s_{00,\nu}^a$ & $a=L$ & $a=R$ & $s_{22,\nu}^a$ & $a=L$ & $a=R$ \\\hline
		$4$ & $1$ & $1$ & $4$ & $1$ & $1$\\
		$5$ & $-1$ & $-1$ & $5$ & $-1$ & $-1$
		\end{tabular}
	\caption{All nonvanishing charge-transfer coefficients $s_{if,\nu}^a$ characterizing the current contributions of individual processes. Sketches of the various processes $\nu=1,\ldots,5$ are depicted in Fig.~\ref{fig:processes}.
	\label{tab}}
\end{table}

It is also possible to obtain a comprehensive analytical understanding of the numerical results based on the Fano factor formula derived in Ref.\ \onlinecite{koch3},
\be\label{fano}
F=\langle N_i \rangle \frac{\langle t_i^2\rangle - \langle t_i \rangle^2}{\langle t_i \rangle^2}+\frac{\langle N_i^2\rangle - \langle N_i \rangle^2}{\langle N_i \rangle}.
\ee
In other words, the Fano factor acquires two distinct contributions: The first term in Eq.~\eqref{fano} arises from the fluctuations of waiting times $t_i$, the second term reflects the fluctuations of the number of transferred electrons $N_i$. Equation \eqref{fano} is valid whenever the quantities $N_i$ and $t_i$ are uncorrelated, and have fixed probability distributions independent of $i$. 
Employing this formula, we analytically calculate the separate Fano factors for cotunneling, unidirectional pair tunneling, and mixed pair tunneling in the following.

\subsection{Cotunneling}
For temperatures small compared to the applied bias, $k_BT \ll eV$, cotunneling transport is essentially unidirectional: transitions with electron transfer in the direction opposed to the applied bias can be neglected. The current is then determined by a single parameter, the cotunneling rate $W$. The evaluation of Eq.~\eqref{fano} is straightforward. The number of transferred electrons per event is exactly one, and the fluctuations of the transmitted charge vanish. The waiting times $t_i$ follow an exponential distribution $p(t)=We^{-W t}$, so that the first and second moments are given by
\be
\langle t_i \rangle = \frac{1}{W},\qquad \langle t_i^2 \rangle = \frac{2}{W^2}.
\ee
As a result, cotunneling leads to a Fano factor of unity,
\be
F= \frac{\langle t_i^2 \rangle-\langle t_i \rangle^2}{\langle t_i \rangle^2}=1.
\ee
This explains the Poissonian noise in the regions dominated by cotunneling in Fig.~\ref{fig:fano}.

\subsection{Unidirectional pair tunneling}
We now consider the case of unidirectional pair tunneling, i.e.\ transport is assumed to be caused by the following sequence of events: (i) An electron pair tunnels from the left lead onto the molecule. (ii) The electron pair tunnels off of the molecule into the right electrode. When evaluating the Fano factor via Eq.~\eqref{fano}, we consider events in one of the two junctions. Without loss of generality, we choose the left junction. Then, each event (i) transfers $N_i=2$ electrons, and there are no fluctuations of this number. However, it is crucial to note that the waiting time $t_i$ now consists of a sum of two times, the waiting time for the event (i) in junction $L$ and the waiting time for event (ii) in junction $R$, $t_i=t_i^L+t_i^R$. Each waiting time $t_i^{L,R}$ is exponentially distributed according to
\begin{align}
p_L(t)&=W_{0\to2}^L \exp \left[-W_{0\to2}^L t\right]\\
p_R(t)&=W_{2\to0}^R \exp \left[-W_{2\to0}^R t\right],
\end{align}
where $W_{0\to2}^L$ ($W_{2\to0}^R$) is the rate for the pair-tunneling transition in the left (right) junction.
The probability distribution $P(t)$ for the sum of  waiting times in the left and right junctions is simply given by the convolution
\begin{align}
P(t)&=\int_0^\infty dt^L \int_0^\infty dt^R\, p_L(t^L) p_R(t^R) \delta(t-t^L-t^R)\nonumber\\
&=\frac{W_{0\to2}^L W_{2\to0}^R}{W_{2\to0}^R-W_{0\to2}^L}\left[e^{-W_{0\to2}^L t}-e^{-W_{2\to0}^R t} \right].
\end{align}
The resulting first and second moments of the total waiting time are
\begin{align}
\langle t_i \rangle &= \frac{1}{W_{0\to2}^L} + \frac{1}{W_{2\to0}^R}=\frac{W_{0\to2}^L+W_{2\to0}^R}{W_{0\to2}^LW_{2\to0}^R}\\
\langle t_i^2 \rangle &= 2\frac{[W_{0\to2}^L]^2+W_{0\to2}^LW_{2\to0}^R+[W_{2\to0}^R]^2}{[W_{0\to2}^L]^2[W_{2\to0}^R]^2}.
\end{align}
Substituting into Eq.~\eqref{fano}, we obtain for the Fano factor
\be
F=\langle N_i \rangle \frac{\langle t_i^2\rangle - \langle t_i \rangle^2}{\langle t_i \rangle^2} = 2\frac{(W_{2\to0}^R)^2+(W_{0\to2}^L)^2}{(W_{2\to0}^R+W_{0\to2}^L)^2}.\label{pairfano1}
\ee
Due to their phase-space behavior, the pair-tunneling rates vary with gate voltage according to
\begin{align}\label{ratebeh1}
W_{0\to2}^L &\sim \frac{\Gamma_L^2}{h}(eV/2 - \varepsilon_d - U/2)\theta(eV/2 - \varepsilon_d - U/2),\\
W_{2\to0}^R &\sim \frac{\Gamma_R^2}{h}(\varepsilon_d + U/2 + eV/2)\theta(\varepsilon_d + U/2 + eV/2),\label{ratebeh2}
\end{align}
valid in the limit of low temperatures.
Consequently, for a symmetric device the rates are identical at the degeneracy point $2\varepsilon_d+U=0$, and the Fano factor resulting from Eq.~\eqref{pairfano1} is $F=1$. Away from the degeneracy point, one rate loses phase space while the other gains phase space. As a result, for approximate alignment of the two-particle level with one of the Fermi energies, $\varepsilon_d+U/2\approx \pm eV/2$, one rate nearly vanishes and the other one remains finite. The corresponding Fano factor is $F=2$. This dependence of the Fano factor on gate voltage is clearly reflected in the numerical results depicted in Fig.~\ref{fig:fano}. Interestingly, for asymmetric devices unidirectional pair-tunneling leads to Fano factors $F>1$ even at the degeneracy point. In the limit $\Gamma_a/\Gamma_{a'}\to\infty$, one obtains a Fano factor of $F=2$. 

It is worth noting that Eq.~\eqref{pairfano1} also yields the Fano factor for sequential tunneling when replacing $\langle N_i \rangle=2$ by $\langle N_i \rangle=1$, as well as $W_{0\to2}^L\to W_{0\to1}^L$ and $W_{2\to0}^R\to W_{1\to0}^R$. The crucial difference between sequential and pair tunneling is that sequential rates are \emph{independent} of gate voltage as long as the level position remains in the bias window. Accordingly, the Fano factor for sequential tunneling in a symmetric device is equal to 1/2.

\subsection{Mixed pair tunneling}
Finally, we consider mixed pair tunneling, i.e.\ the transport mode typical for pair tunneling in asymmetric junctions.  Two electrons enter the molecule from the left lead (rate $W_1$), and subsequently split up to leave the molecule via the left and right junction, respectively (rate $W_2$). 
Let us consider the right junction. Here, exactly one electron is transferred for each combined process of pair tunneling onto and off of the molecule. The waiting time again consists of a sum of two exponentially distributed random times, exactly as in the unidirectional  pair-tunneling case. All arguments given there can be reapplied for the case of the mixed pair-tunneling case, resulting in
\be
F=\frac{W_1^2+W_2^2}{(W_1+W_2)^2},
\ee
Formally, this is identical to the result for sequential tunneling. Once again, the important difference is the gate-voltage dependence of the pair-tunneling rates. While the phase-space behavior of the rate $W_1$ is given by Eq.~\eqref{ratebeh1}, the splitting pair rate scales like 
\be
W_2\sim\frac{\Gamma_L\Gamma_R}{h}(2\varepsilon_d+U)\theta(2\varepsilon_d+U).
\ee Consequently, the Fano factor for the mixed pair-tunneling process varies between $F=1$ and $F=1/2$, depending on the coupling ratio and gate voltage.

\subsection{Interpretation of the full Fano factor}
The full Fano factor, Fig.~\ref{fig:fano}, can now easily be interpreted in terms of the contributions from the previously discussed processes. Outside the pair-tunneling regime, cotunneling is the only relevant process and leads to a Fano factor equal to 1. Inside the pair-tunneling regime, cotunneling, as well as unidirectional and mixed pair tunneling \emph{coexist}. The resulting Fano factor is given by a weighted average of the individual Fano factors. This comprehensively explains the qualitative features of the numerical results in Fig.~\ref{fig:fano}, and the fact that the Fano factor does not fully reach the value of 2 naively expected for pair tunneling.
\begin{figure}
	\centering
	\psfrag{x}[][][0.9]{\raisebox{0.1cm}{$\varepsilon_d+\abs{U}/2$ ($10^{-3}\abs{U}$)}}
	\psfrag{y}[][][0.9]{\raisebox{0.1cm}{$F$}}
	\psfrag{a}[l][][0.9]{\raisebox{0.1cm}{$\Gamma_R/\Gamma_L=1$}}
  \psfrag{b}[l][][0.9]{\raisebox{0.1cm}{$\Gamma_R/\Gamma_L=2.25$}}
  \psfrag{c}[l][][0.9]{\raisebox{0.1cm}{$\Gamma_R/\Gamma_L=25$}}
  \psfrag{d}[l][][0.9]{\raisebox{0.1cm}{$\Gamma_R/\Gamma_L=100$}}
		\includegraphics[width=0.95\columnwidth]{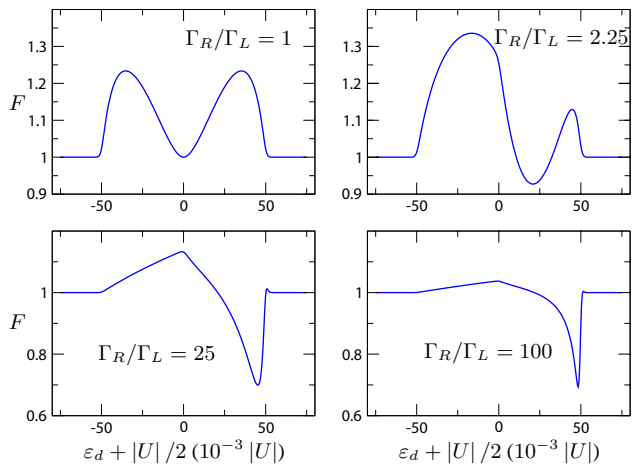}
	\caption{Fano factor as a function of gate voltage at finite bias ($eV=50\times10^{-3}\abs{U}$) for different magnitudes of junction asymmetry $\Gamma_L/\Gamma_R$. The dominance of mixed pair-tunneling processes for asymmetric junctions reduces the super-Poissonian noise and leads to Fano factors below 1.\label{fig:fano-2}}
\end{figure}

For completeness, we briefly comment on the consequences of asymmetric coupling $\Gamma_L\not=\Gamma_R$ on the shot noise in the negative-$U$ model. As discussed for current-voltage characteristics in Ref.\ \onlinecite{koch6}, the degree of coupling asymmetry plays an important role in determining the relevant transport processes. As a result, the noise characteristics crucially depend on the ratio $\Gamma_R/\Gamma_L$, see Fig.\ \ref{fig:fano-2}. These plots show results for the finite-bias Fano factor as a function of gate voltage, and for different coupling ratios. 
For devices with large asymmetry, $\Gamma_{a}/\Gamma_{a'}\gg1$, unidirectional pair-tunneling is suppressed and mixed pair processes take over.\cite{koch6} As a result, the Fano factor is reduced, and sub-Poissonian noise dominates the transport, see Fig.~\ref{fig:fano-2}. The influence of unidirectional pair-tunneling close to the degeneracy point is reflected in remaining super-Poissonian traces for moderate asymmetries.


\begin{thebibliography}{30}
\bibitem[{fno()}]{fno1}
\bibinfo{journal}{When using this notation (as opposed to the two-particle
  energy $2\varepsilon_d+U$), the pair level can be accessed in a real process
  from the left lead whenever it is situated below the left Fermi level.}
\end{thebibliography}
\end{document}